\documentclass[12pt,a4paper]{article}
\usepackage{graphicx}
\usepackage{amssymb}
\usepackage{amsthm}
\usepackage{amsmath}
\usepackage{amsfonts}
\usepackage{algorithm, algorithmic}
\usepackage{color}

\usepackage{lineno}

\usepackage{enumitem}

\newtheorem{thm}{Theorem}[section]

\theoremstyle{definition}

\theoremstyle{remark}
\newtheorem{rem}{Remark}[section]

\numberwithin{equation}{section}

\newcommand{\thmref}[1]{Theorem~\ref{#1}}

\begin{document}


\title{In-situ associative permuting}
\author{A. Emre CETIN}

\maketitle

\begin{abstract}

The technique of in-situ associative permuting is introduced which is an association of in-situ permuting and in-situ inverting. It is suitable for associatively permutable permutations of $\lbrace 1,2,\ldots,n \rbrace$ where the elements that will be inverted are negative and stored in order relative to each other according to their absolute values.

Let $K[1\ldots n]$ be an array of $n$ integer keys each in the range $[1,n]$, and it is allowed to modify the keys in the range $[-n,n]$. If the integer keys are rearranged such that one of each distinct key having the value $i$ is moved to the $i$th position of $K$, then the resulting arrangement (will be denoted by $K^P$) can be transformed in-situ into associatively permutable permutation $\pi^P$ using only $\log n$ additional bits. The associatively permutable permutation $\pi^P$ not only stores the ranks of the keys of $K^P$ but also uniquely represents $K^P$. Restoring the keys from $\pi^P$ is not considered. However, in-situ associative permuting $\pi^P$ in $\mathcal{O}(n)$ time using $\log n$ additional bits rearranges the elements of $\pi^P$ in order, as well as lets to restore the keys of $K^P$ in $\mathcal{O}(n)$ further time using the inverses of the negative ranks. This means that an array of $n$ integer keys each in the range $[1,n]$ can be sorted using only $\log n$ bits of additional space.

\end{abstract}




\section{Introduction}\label{sec:intro_abs}

A solution for sorting an array $K[1\ldots n]$ of $n$ integer keys is to find a permutation $\pi^-_1\pi^-_2\ldots\pi^-_n$ of the indices such that, $K_{\pi^-_1} \le K_{\pi^-_2} \le \ldots \le K_{\pi^-_n}$. Once $\pi^-$ is found, it is sufficient to access the keys in order of their ranks. This is address table sort~\cite{knuth:vol3}. 

Another solution is the inverse of $\pi^-$. Each $\pi_i$ is the rank of the key $K_i$, describing where it should be placed when the keys are rearranged in order of their ranks. Comparison counting sort~\cite{knuth:vol3} finds $\pi$ in $\mathcal{O}(n^2)$ time.



Once $\pi$ (or $\pi^-$) is found, its inverse $\pi^-$ (or $\pi$) can be found in-situ in $\mathcal{O}(n)$ time using $n$ additional bits. If additional $n$ bits are not allowed, it is possible to tag the elements by making them negative when they are inverted~\cite{knuth:vol1}. At the end, the keys can be restored by correcting the signs. 


In some situations, the integer keys need to be stored in order of their ranks. If $\pi$ or $\pi^-$ is given in an array and it is allowed to modify the given, the rearrangement can be done in-situ in $\mathcal{O}(n)$ time and this is known as in-situ permuting (cycle leader permutation)~\cite{knuth:vol3,knuth:vol1}. 

Sometimes, it may not be possible to modify $\pi$. This is firstly investigated by~\cite{knuth_1}, and then by~\cite{Fich}. In summary, if it is not allowed to use additional space to tag the element, time complexity of in-situ permuting is $\mathcal{O}(n \log n)$.

If the cumulative distribution is computed in an array $\lambda$, then $\pi$ can be computed explicitly from $\lambda$. However, considering the space requirements, it is reasonable that the efforts are on searching techniques for (i) classifying the keys to reduce $\lambda$, (ii) in-situ permuting the keys by computing the ranks implicitly from $\lambda$ as an oracle $\pi(i)$. Distribution counting sort~\cite{knuth:vol3}, address calculation sort~\cite{Isaac} and bucket sort\cite{mahmoud:2000} are some important examples. 


In Section~\ref{ap}, the technique of in-situ associative permuting will be introduced which is an association of in-situ permuting and inversion. Then, in Section~\ref{stable_dcs}, in-situ associative permuting sort will be introduced which will be followed by the conclusions.

\section{Associatively Permutable Permutations}\label{ap}

{\bf Problem statement:} Let $\pi_1,\pi_2,\ldots,\pi_n$ is an associatively permutable permutation of $\lbrace 1,2,\ldots,n \rbrace$ where the elements that need to be inverted are negative and in order relative to each other according to their absolute values. The problem is in-situ inverting of the negative elements while the positive ones are in-situ permuted. The signs of the elements should remain the same. In first step, the solution will be given which will be followed by the proof.



{\bf In-situ associative permuting:} Starting with the first positive rank, an outer cycle leader permutation can move only the positive elements to their final position ignoring the negative ones. This is possible since when a positive element having the value $i$ is moved to its final position $\pi_i$, it will be tagged by $\pi_i = i$. If a positive index is moved onto a negative index, then until a positive index is encountered again, an inner cycle leader permutation can move only the negative elements to their final position storing negative of their former position, which is the same with inverting the negative elements. When a positive index is encountered again, the inner loop can stop and the outer loop can continue until all the positive elements are in-situ permuted.


\begin{thm}\label{thm1}
Given an associatively permutable permutation $\pi[1 \ldots n]$ in which the negative indices are in order relative to each other (according to their absolute values), the above $\mathcal{O}(n)$ time solution can in-situ permute $\pi$ associatively.
\end{thm}

\begin{proof} If there are $1 \le n_d \le n$ negative indices, any negative index $\pi_i$ is either a singleton cycle ($\pi_i = -i$) or a part of another disjoint cycle. If it is a singleton cycle, then its inverse is already equal to itself and there are not any positive or negative indices which will be moved to the place of $\pi_i$. On the other hand, if it is a part of a disjoint cycle, the only case in which an inner cycle leader permutation can not be started on $\pi_i$ is when there are not any positive indices involved in that particular disjoint cycle. Let two negative indices $\pi_i$ and $\pi_j$ form a disjoint cycle $(\pi_i\pi_j)$ without a positive index. This means $\pi_i$ and $\pi_j$ address each other by $\pi_i =-j$ and $\pi_j =-i$. However, this contradicts with the assumption that the negative indices are in order relative to each other according to their absolute values. In other words, if $\vert \pi_i \vert < \vert \pi_j \vert$, then $\vert \pi_i \vert = j$ implies $\vert \pi_j \vert  > j$ or $\vert \pi_j \vert = i$ implies $\vert \pi_i \vert  < i$. Therefore, there exists at least one positive index in every disjoint cycle which includes at least one negative index. On the other hand, if there are not any positive indices in $\pi$ or for $1 \le r < n$, all possible $r$-combinations form a disjoint cycle (singleton cycles are indeed disjoint cycles), then there is only one arrangement for relatively ordering $n_d$ negative indices in $n_d$ places according to their absolute values, which implies that each negative index is indeed a singleton cycle and hence its inverse is equal to itself.
\end{proof}

\begin{thm}\label{thm2}
Given an array $K[1\ldots n]$ of $n$ integer keys each in the range $[1,n]$ and the permutation $\pi[1\ldots n]$ of the indices $\lbrace 1,2,\ldots,n \rbrace$ corresponding to the ranks of the keys, 
\begin{enumerate}[label=(\roman{*}).]
 \item if each minimum-ranked distinct key is tagged by making its rank negative in $\pi$, and,

\item considering $\pi$ as the records of the keys, if $\pi$ is rearranged according the keys, such that each negatively ranked key having the value $i$ in $K$ is moved to the $i$th position of $K$,

\end{enumerate}
then the resulting $\pi$ (hereafter will be denoted as $\pi^P$) is in-situ associatively permutable and uniquely represents the resulting $K$ (hereafter will be denoted as $K^P$).

\end{thm}

\begin{proof}

After each minimum-ranked distinct key has been tagged by making its rank negative in $\pi$ somehow, the rearrangement can be done in-situ in $\mathcal{O}(n)$ time since each relocated key will be tagged by $K^P_i = i$. After the rearrangement, not only does each negative rank $\pi^P_i$ describe the negative of the position (rank) at which the key $K^P_i$ will be stored when the keys are rearranged in order of their ranks, but also registers the value $i$ of the key in its index. Moreover, the rearrangement ensures that $\pi^P$ is in-situ associatively permutable, i.e., the negative ranks are stored in order relative to each other in $\pi^P$ according to their absolute value, since given two negative ranks $\vert \pi^P_i \vert$, $\vert \pi^P_j \vert$, $i < j$ implies $K^P_i < K^P_j$ and hence $\vert \pi^P_i \vert < \vert \pi^P_j \vert$. Hence, from left to right, if $\pi^P_i$ is the first and $\pi^P_j$ is the second negative rank, then the positive ranks in $\pi^P$ having the values $ \vert \pi^P_i \vert + 1,\ldots,\vert \pi^P_j \vert -1 $ together with $\pi^P_i$ are the corresponding keys of $K^P$ that are all equal to $i$. Besides, if $\pi^P_j$ is the last negative rank, then the positive ranks in $\pi^P$ having the values $ \vert \pi^P_j \vert + 1,\ldots,n $ together with $\pi^P_j$ are the corresponding keys of $K^P$ that are all equal to $j$. 
\end{proof}

Restoring the keys back from $\pi^P$ is out of the scope of this study. However, it is important to notice that, for every consecutive negative ranks $\pi^P_i$ and $\pi^P_j$, if $\vert \pi^P_j - \pi^P_i \vert > 1$, then in worst case, $\mathcal{O}(n)$ time is required to find in $\pi^P$ the $(\vert \pi^P_j - \pi^P_i \vert-1)$ positive ranks having the values $\vert \pi^P_i \vert + 1,\ldots,\vert \pi^P_j \vert -1$ and restore them. On the other hand, \thmref{thm1} and \thmref{thm2} let us to assert that:
\begin{thm}\label{thm3}
In-situ permuting $\pi^P$ associatively in $\mathcal{O}(n)$ time rearranges the ranks in order, as well as lets to restore the keys in $\mathcal{O}(n)$ further time by using the inverses of the negative ranks.  
\end{thm}

\begin{proof}
Before in-situ permuting $\pi^P$ associatively, not only does each negative rank $\pi^P_i$ describe the negative of the position (rank) at which the key $K^P_i$ will be stored after the rearrangement, but also registers the value $i$ of the key in its index. Hence, while in-situ permuting $\pi^P$ associatively, each negative rank $\pi^P_i$ is moved to its final position $\vert \pi^P_i \vert$ storing the negative value of the corresponding key by $\pi^P[\vert \pi^P_i \vert] \leftarrow -i$.  At the end, each inverse $\pi^P_i$ precedes the positive ranks of the keys having the value $i$ (i.e., $\vert \pi^P_i \vert$) until the next inverse. In other words, if $\pi^P_i$ is the first and $\pi^P_j$ is the second inverse, then the positive ranks $\pi^P_{i+1},\ldots , \pi^P_{j-1}$ having the values $i+1,\ldots,j-1$ together with the inverse $\pi^P_i$ are the corresponding keys of $K^P$ that are all equal to $\vert \pi^P_i \vert$. Furthermore, if $\pi^P_j$ is the last inverse of $\pi^P$, then the positive ranks $\pi^P_{j+1},\ldots, \pi^P_n$ having the values $ j+1,\ldots,n$ together with the inverse $\pi^P_j$ are the corresponding keys of $K^P$ that are all equal to $\vert \pi^P_j \vert$.

\end{proof}

\section{In-situ Associative Permuting Sort}\label{stable_dcs}

Consider that all the positive ranks of $\pi^P$ are set to zero without loosing generality assuming equal keys are resolved arbitrarily. In such a case, each nonzero $\pi^P_i$ stores the negative of the position (rank) at which the key $K^P_i$ having the value $i$ will be stored when the keys are rearranged in order of their ranks. On the other hand, if one counts, towards negative, the number of repeating keys of $K^P$ in an array $\lambda^P[1 \ldots n]$ and computes a prefix sum on only its nonzero elements starting the accumulation from zero, each nonzero $\lambda^P_i$ stores the negative of the last position (rank) at which the key $K^P_i$ having the value $i$ will be stored when the keys are rearranged in order of their ranks. Furthermore, remembering that the elements of $\lambda^P$ and $\pi^P$ are negative, if one processes the keys of $K^P$ for $i=n,n-1,\ldots,1$, decreasing $\pi^P[K^P_i]$ and increasing $\lambda^P[K^P_i]$ whenever $K^P_i \ne i$, then at the end $\pi^P$ becomes $\lambda^P$, whereas $\lambda^P$ becomes $\pi^P$. It is immediately noticed that the keys that satisfy $K^P_i = i$ are not processed. In other words, if it is allowed to modify the keys of $K^P$ in the range $[-n,n]$, then one can modify the keys that satisfy $K^P_i=i$ making them $K^P_i\leftarrow -1$, and another can restore the keys that satisfy $K^P_i<0$ making them $K^P_i\leftarrow i$. This lets us to assert that:


\begin{thm}\label{thm4}
Given an array $K[1\ldots n]$ of $n$ integer keys each in the range $[1,n]$, if it is allowed to modify the keys in the range $[-n,n]$, then $K$ can be transformed into $\pi^P$ that uniquely represents $K^P$ using only $\log n$ additional bits. 
\end{thm}

\begin{proof} $K$ can be transformed into $\pi^P$ by carrying out the following tasks:
\begin{enumerate}[label=(\roman{*}).]
\item In-situ rearrange $K$ such that one of each distinct key having the value $i$ is moved to the $i$th position. This is possible using only additional $\log n$ bits since each relocated key is tagged by $K_i=i$. 
\item Set all the keys that satisfy $K_i =i$ to $K_i\leftarrow -1$. 
\item Whenever $K_i \ge 0$, decrease $K[K_i]$ by one, for $i=1,2,\ldots, n$. 
\item Compute a prefix sum on negative elements of $K$. After the prefix sum, the negative elements of $K$ represent $\lambda^P$.
\item Whenever $K_i \ge 0$, increase $K[K_i]$ by one and set $K_i \leftarrow -K[K_i] + 1$, for $i=n,n-1,\ldots 1$. Direction is not important since we sacrificed stability in the first step. At the end, $K$ becomes the associatively permutable permutation $\pi^P$.

\end{enumerate}
\end{proof}

\begin{rem}
It should be noted that, the first three steps can be combined within one loop for $i=1,2,\ldots,n$. 
\end{rem}

Once $\pi^P$ is obtained, it can be in-situ permuted associatively in $\mathcal{O}(n)$ time which puts the ranks in order, as well as lets to restore the keys in $\mathcal{O}(n)$ further time by using the inverses of the negative ranks (\thmref{thm3}).

\section{Conclusions}\label{chap:summaryandconclusion}
The technique of in-situ associative permuting has been introduced which is an association of in-situ permuting and in-situ inverting. It is suitable for associatively permutable permutations $\pi^P$ where the elements that need to be inverted are negative and stored in order relative to each other according to their absolute values. 

Given an array $K$ of $n$ integer keys each in $[1,n]$, and the permutation $\pi$ of the indices corresponding to the ranks of the keys, $K^P$ and associatively permutable permutation $\pi^P$ that uniquely represents $K^P$, can be obtained from $\pi$ and $K$. This means that the memory allocated for $K^P$ is gained. Restoring $K^P$ from $\pi^P$ is not considered in this study. However, in-situ associative permuting $\pi^P$ puts the ranks (hence the keys) in order, as well as lets to restore the keys in further $\mathcal{O}(n)$ time by using the inverses of the negative ranks.

The transformations between $K$, $\pi$ and $\lambda$ have been studied. Although, there are certainly others, one of them has been introduced which first rearranges $K$ unstably into $K^P$ and than transforms it into associatively permutable permutation $\pi^P$ which uniquely represents $K^P$. If it was possible to in-situ rearrange $K$ into $K^P$ in a stable manner and all the most significant bits of the integer keys were empty, i.e., each in $[1,2^{w-1}]$ where $w$ is the fixed word length, then in-situ associative least significant $\lceil \log n \rceil$-base radix permuting sort would be possible. On the other hand, if each integer key is in $[1,2^{w-1}]$, instable in-situ associative most significant $\lceil \log n \rceil$-base radix permuting sort has been developed which requires only $\log n$ additional bits. It has been tested with positive results; up to $2^{20}$ integer keys each in $[1,2^{20}]$, radix sort that use $n$ additional words is roughly two times faster, whereas it is faster roughly $1.5$ times than quick sort which uses $\log n$ additional words. On the other hand, when $m < \frac{n}{10}$, it outperforms radix sort. 





\begin{thebibliography}{00}


\bibitem{knuth:vol3} {D.E. Knuth, The Art of Computer Programming, Volume 1: Fundamental Algorithms, Addison-Wesley, 1997.}
\bibitem{knuth:vol1} {D.E. Knuth, The Art of Computer Programming, Volume 3: Sorting and Searching, Addison-Wesley, 1998.}
\bibitem{knuth_1} {D.E. Knuth, ``Mathematical analysis of algorithms'', Proc. of IFIP Congress, pp. 19-–27, 1971.}
\bibitem{Fich} {F.E. Fich, J.I. Munro, P.V. Poblete, ``Permuting in-place'', SIAM J. Comput., Vol. 24, pp. 266~–~278, 2006.}








\bibitem{Isaac} {E.J. Isaac, R.C. Singleton, ``Sorting by address calculation'', J. of ACM,  Vol. 3, pp. 169--174, 1956.}



\bibitem{mahmoud:2000} {H.M. Mahmoud, Sorting, A Distribution Theory, John Wiley and Sons, 2000.}

\bibitem{rosen:handbook}{H.K. Rosen, Handbook of Discrete and Combinatorial Mathematics, CRC Press, 2000}

\end{thebibliography}


\end{document}